%% file: main.tex
\documentclass[xcolor,pdftex,usenames,dvipsnames,table]{PoS}

\usepackage{multirow, lscape}
\usepackage{slashed}
\usepackage{graphicx}
\usepackage{amssymb,amsmath}
\usepackage{url}

\title{Is there any gender/race bias in hep-lat primary publication? Machine-Learning Evaluation of Author Ethnicity and Gender}

\ShortTitle{Gender/Race Bias Study using Machine Learning Evaluation} 

\author{Huey-Wen Lin\\
        Department of Physics and Astronomy,
        Michigan State University, MI, 48824, U.S.A\\
        Department of Computational Mathematics,
        Science and Engineering, Michigan State University, MI, 48824, U.S.A
         E-mail: \email{hwlin@pa.msu.edu}
        }

\abstract{
In this work, we analyze papers that are classified as primary hep-lat to study whether there is any race or gender bias in the journal-publication process.
We implement machine learning to predict the race and gender of authors based on their names and look for measurable differences between publication outcomes based on author classification.
We would like to invite discussion on how journals can make improvements in their editorial process and how institutions or grant offices should account for these publication differences in gender and race.
}

\FullConference{The 38th Annual International Symposium on Lattice Field Theory - LATTICE2021\\
		26-30 July, 2021\\
		Massachusetts Institute of Technology, Cambridge, Massachusetts, USA.}
		
\maketitle

\begin{document}
\section{Introduction}

There is a slow but growing effort to improve the diversity of STEM fields.
In May 2020, the National Science Foundation (NSF) Directorate for Mathematical and Physical Sciences (MPS) encouraged Principal Investigators (PIs) in an open letter to improve diversity and retention at the doctoral level within MPS~\cite{NSF}. 
In December 2020, the nuclear-physics office of thee U.S. Department of Energy (DOE) announced a pilot program to provide \$3 million for research traineeships to broaden and diversify the nuclear-physics research community~\cite{DOE}. 
Many higher-education institutions are also investing in outreach efforts targeting the pipeline issue in STEM by funding REU programs and K-12 activities locally throughout the year. 
All these efforts are designed to create a wave of new-generation scientists that are more diverse than ever, and ready to tackle hard-to-solve problems by bringing with them their unique way of thinking and problem-solving ability. 
In the late stages of the pipeline within academic jobs, scientists are subjected to the evaluation based on their publications.
This affects the ability of students to get postdoc jobs, of postdocs to get faculty positions, of faculty to get grants and achieve tenure, and so on.
At the highest level, publications are necessary to getting grants, finding permanent positions, getting promotions, and being considered for an awards.
Any significant gender or racial bias~\cite{InequalityPublishing}
could cause years of effort bringing up a diverse workforce to be in vain.
Therefore, it is timely, when the historical minority numbers begin to improve, to study and monitor any potential bias and make sure that there is no bottleneck in these final stages.
In this article, we consider all the papers that have been classified as primarily hep-lat to try to find the current baseline, so that we can revisit these numbers in 5--10 years.

We invite readers to consider these questions:
If there already exist signs of bias, what is its significance? 
What can journals do about it?
How should institutions or grant offices account for these differences, if there is no foreseeable improvement by journals? 

\section{Data Collection}

\begin{figure}[tb]
\centering
\includegraphics[width=0.5\textwidth]{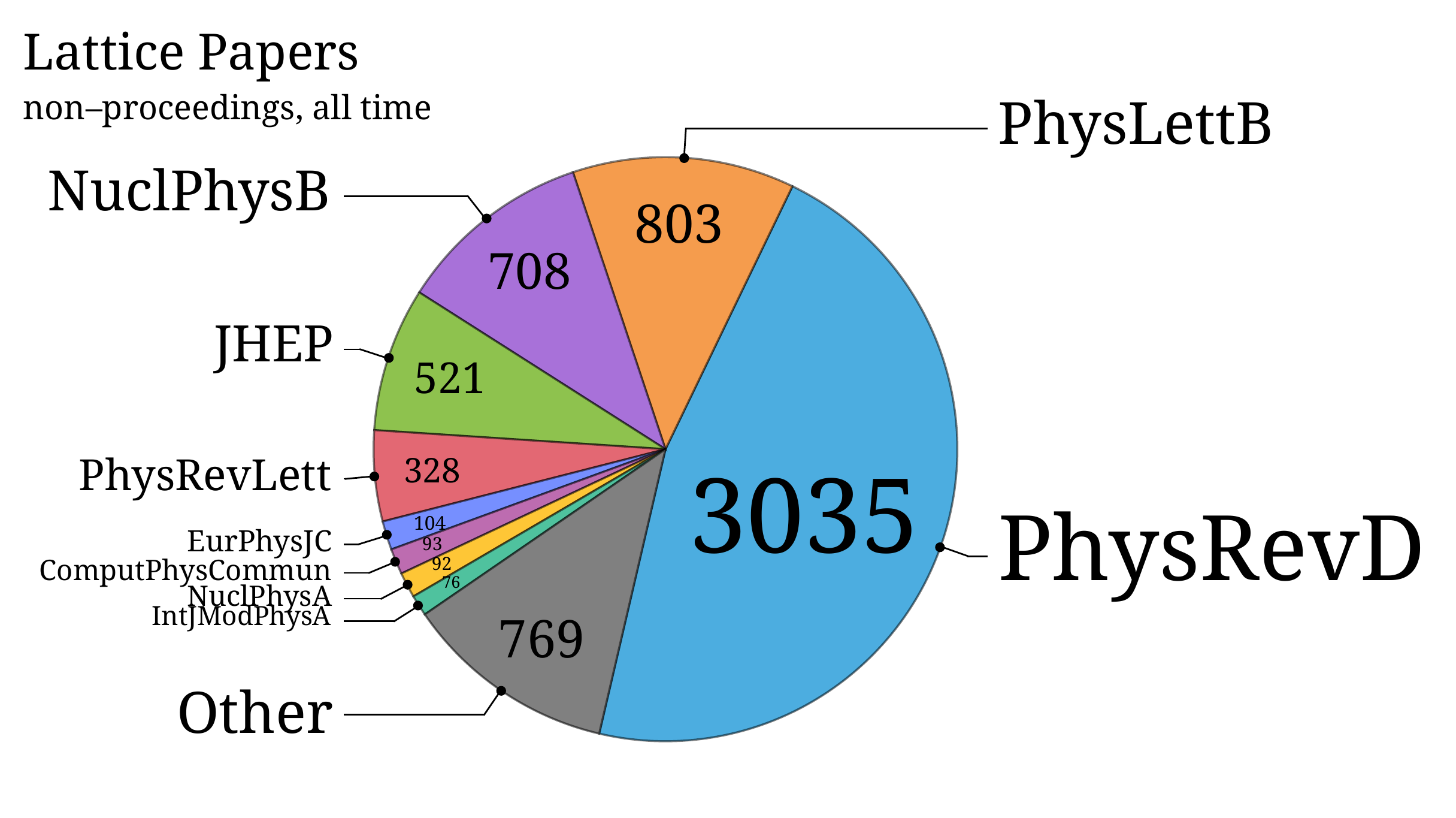}
\includegraphics[width=0.4\textwidth]{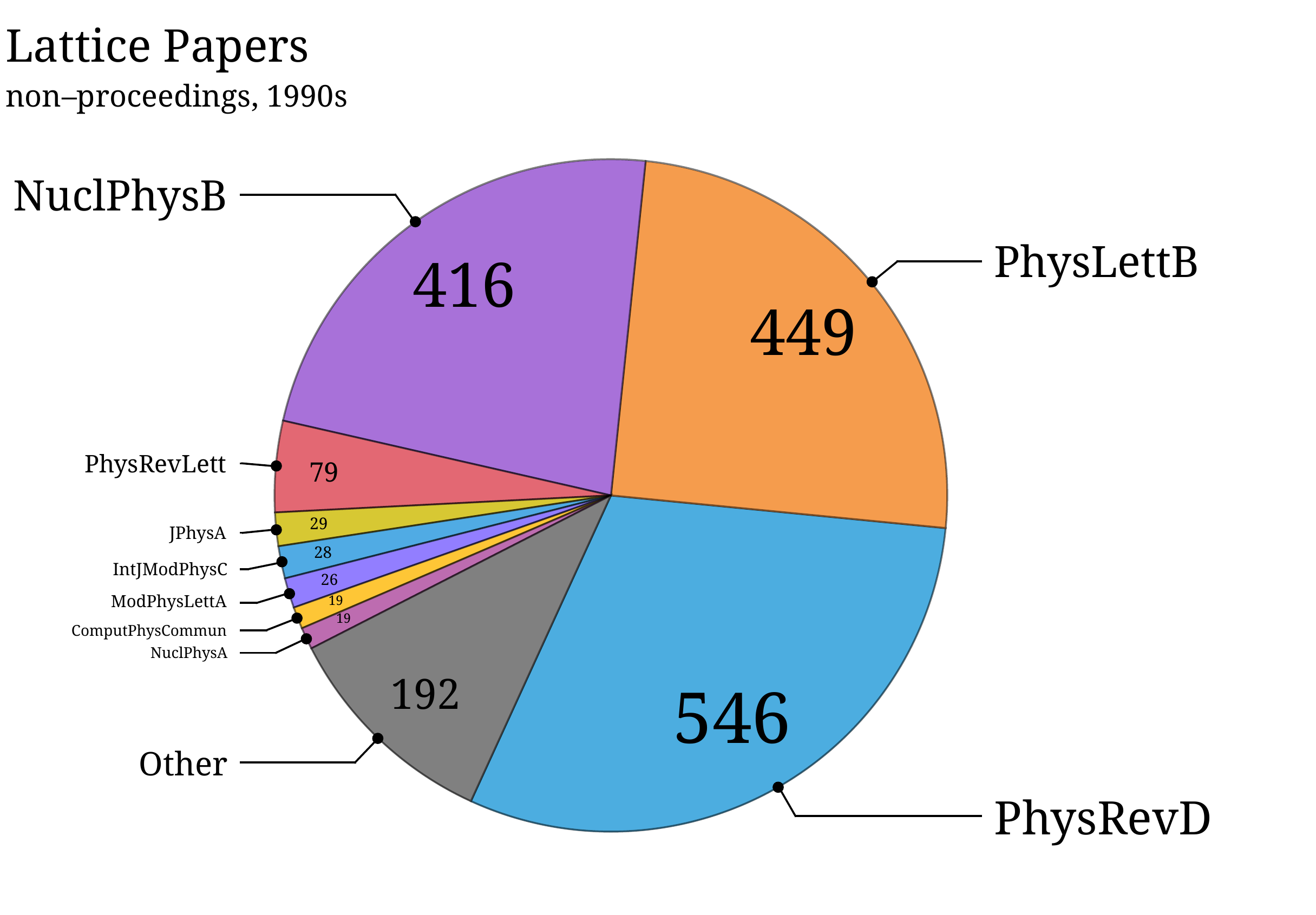}
\includegraphics[width=0.4\textwidth]{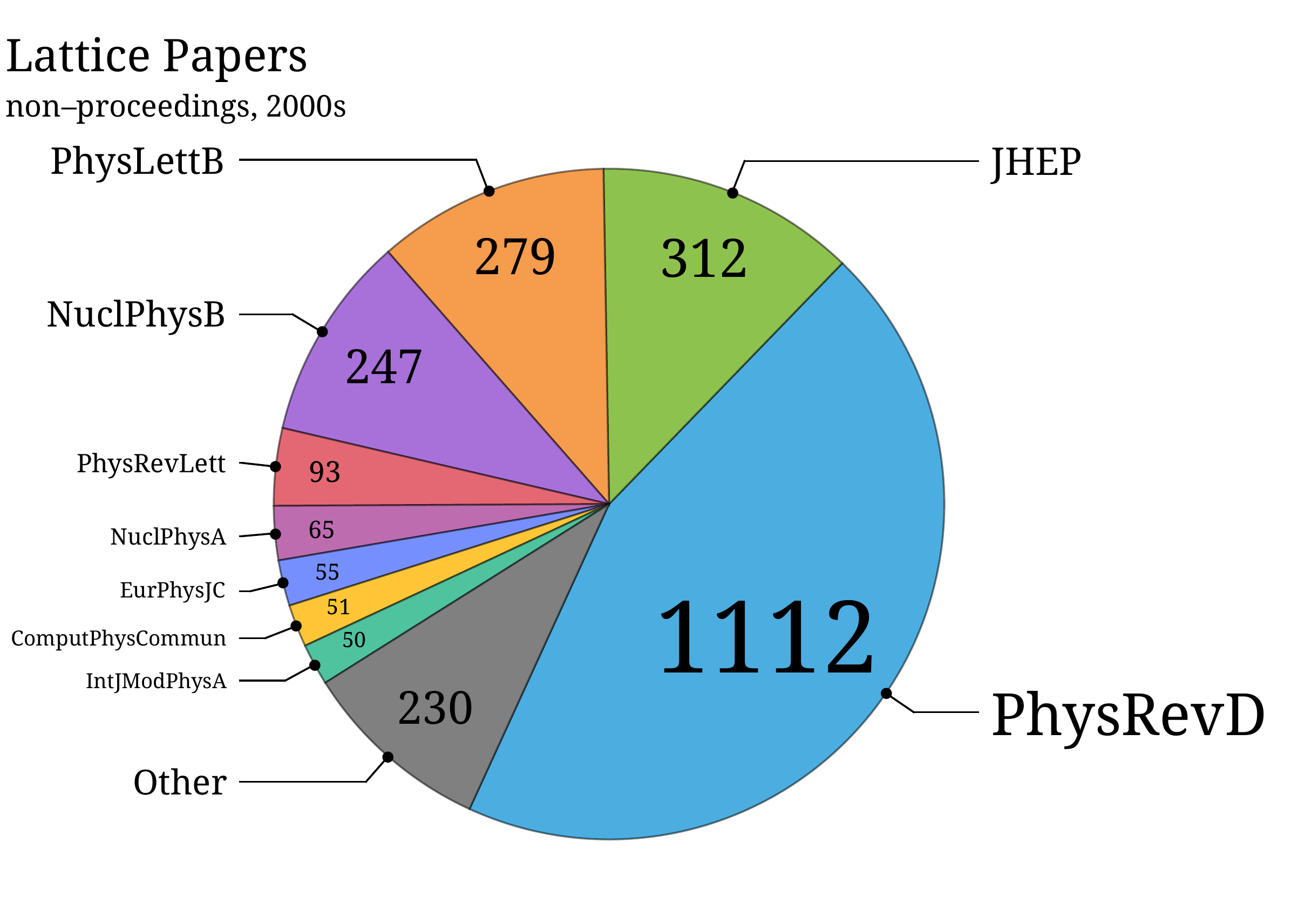}
\includegraphics[width=0.4\textwidth]{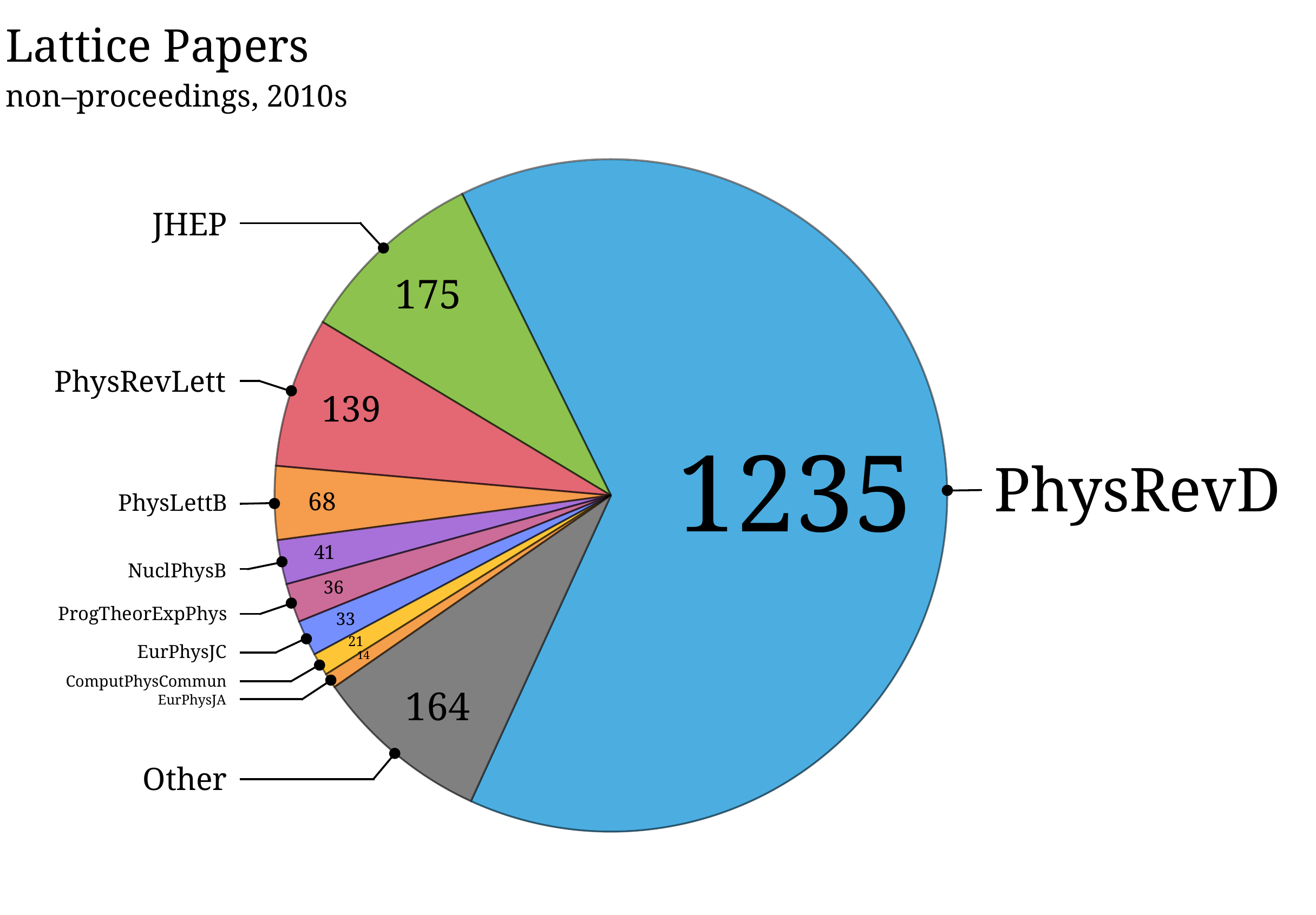}
\caption{
The non-proceeding lattice publications pie charts (upper-left) since the beginning of the arXiv and (others) divided by decade.
\label{fig:journals-pie-by-decades}}
\end{figure}

The first step toward further analysis is to collect the available data and prepare it for processing.
We downloaded the full metadata of all hep-lat papers from the arXiv using their API, and we removed all the hep-lat cross-listed papers and proceedings (for example, from \textit{Nuclear Physics Proceedings Supplement}, \textit{Proceedings of Science}, \textit{Journal of Physics Conference Series}, etc.) that is not relevant to the purpose of this study.
Figure~\ref{fig:journals-pie-by-decades} shows pie charts of which journals lattice papers have been published in since the beginning of the arXiv.
There are a few interesting facts:
the fraction of papers published in \textit{Physical Review~D} (PhysRevD) has steadily increased over the past decades, while papers published in \textit{Physics Letters~B} (PhysLettB) and \textit{Nuclear Physics~B} (NuclPhysB) have diminished.
The fraction of lattice papers published in the \textit{Journal of High-Energy Physics} (JHEP) peaked in the 2000s but declined in the 2010s.
Lattice papers published in \textit{Physical Review Letters} (PhysRevLett) have significantly increased in recent years, likely correlated with the emergence of realistic calculations at the physical pion mass in recent years, which are of broad interest, since they are able to contribute significantly to many experimental quantities.

A major opportunity for race and gender bias to enter the system is for papers to simply be turned away by a journal.
Unfortunately, such interactions are not reflected in the available data and are, thus, impossible for us to analyze.
However, another metric which may show evidence of bias is the delay of a paper's publication.
This is not trivial to determine either, since not all journals report the date of submission of a manuscript.
We can, however, estimate it by examining the time elapsed between the appearance of a preprint on the arXiv and the publication of that paper in a journal.

\begin{figure}[tb]
\centering
\includegraphics[width=0.36\textwidth]{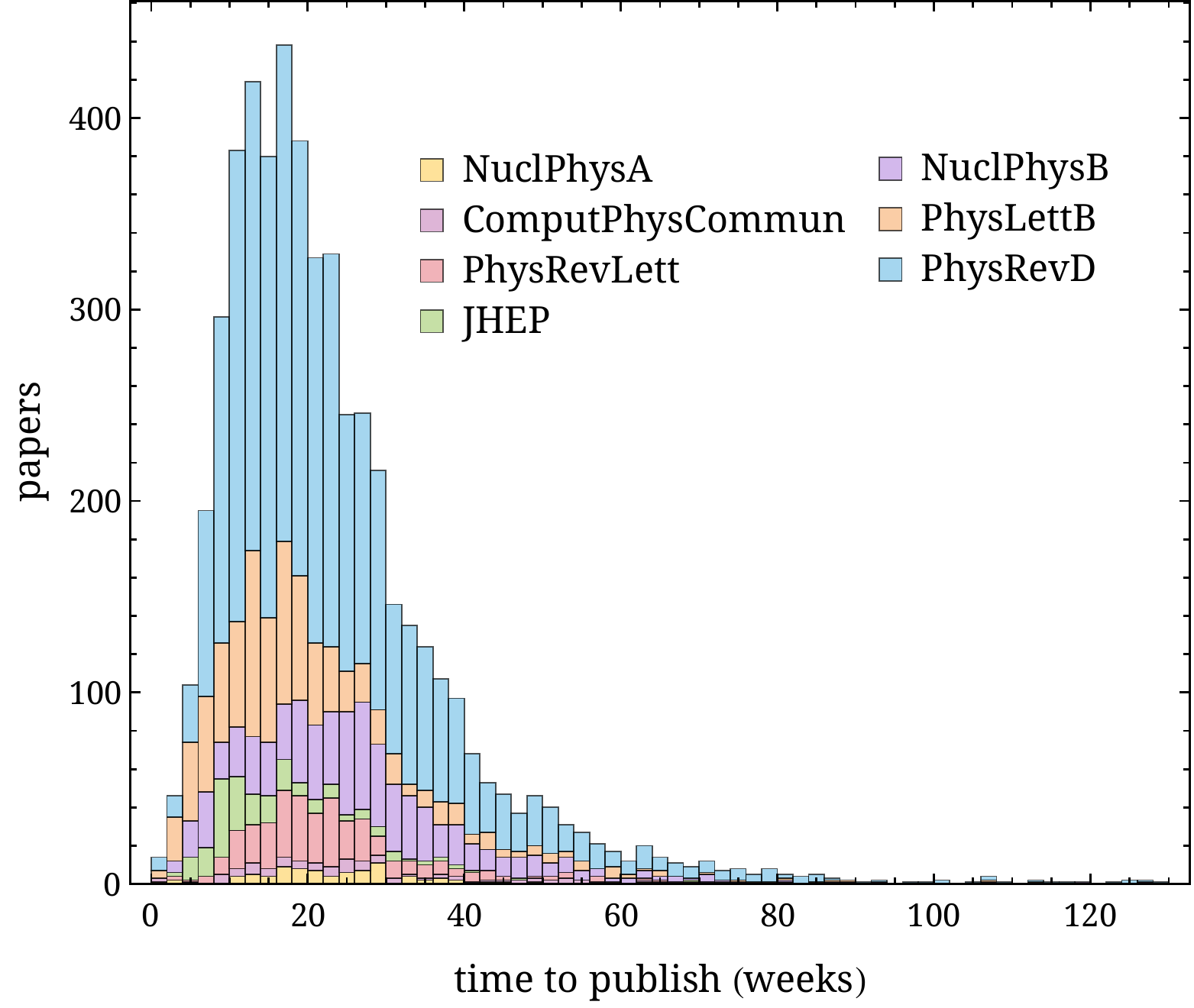}
\includegraphics[width=0.59\textwidth]{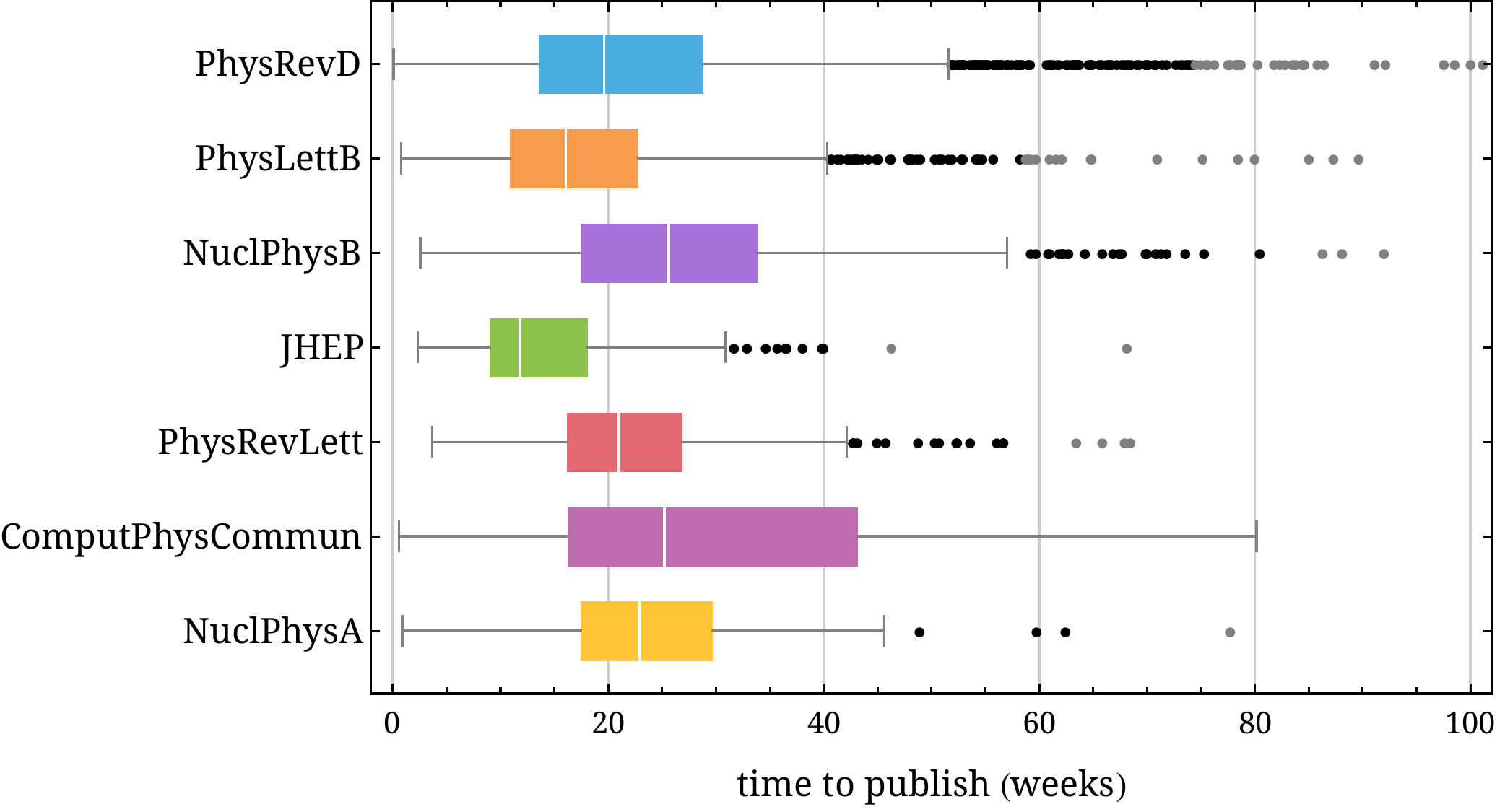}
\caption{
(left) Histogram of the time in weeks between the arXiv and publication dates of non-proceeding papers in top lattice journals. (right) Box plots for the same data.
\label{fig:TopJounalPublicationTime}}
\end{figure}

Using the arXiv metadata, we extract the DOIs for each publication.
By visiting each paper's webpage, we can extract the publication date if it appears on the page.
The arXiv date of the paper's preprint is available directly from the metadata, and we use the difference as the time it took for the paper to published.
Since each publisher uses a different webpage layout, we write a custom scraper for the top publishers:
\begin{itemize}
\item the American Physical Society (APS): PhysRevD, PhysRevLett, etc.
\item Elsevier: PhysLettB, NuclPhysB, \textit{Computer Physics Communications} (ComputPhysCommun), \textit{Nuclear Physics~A} (NuclPhysA), etc.
\item Institute of Physics (IOP): JHEP (--2007), \textit{Journal of Physics~A} (JPhysA), etc.
\item Springer: JHEP (2007--), \textit{European Physics Journal~C} (EurPhysJC), etc.
\item World Scientific: \textit{International Journal of Modern Physics~A} (IntJModPhysA), \textit{Modern Physics Letters~A} (ModPhysLettA), IntJModPhysC, etc.
\end{itemize}
Unfortunately, this is not always possible with automated tools;
IOP swiftly blacklisted our tool, so JHEP publication data is only available after 2007, when it switched to Springer publishing.
Figure~\ref{fig:TopJounalPublicationTime} shows (left) a histogram as a function of time to publish for the top journals.
On the right-hand side, we show box plots (right) of the publication time so that the median and interquartile range is more easily distinguished.
We can see that the distribution is quite long-tailed, with a median above 20 weeks for most journals, excepting JHEP (12 weeks) and PLB (16 weeks).

\section{Automatically Assigning Author Ethnicity Using Machine Learning}

In order to determine the ethnicity of each of the thousands of authors in the dataset, we turn to machine learning.
By training a neural network (NN) on surnames, we can use machine learning (ML) to automatically assign a predicted ethnicity to each author.
We use Jupyter notebooks in Python~3 with NumPy and TensorFlow to perform the NN training;
The TensorFlow backend leverages GPU acceleration via NVIDIA's cuDNN deep-neural-network library.

Our first choice of training data is the United States Census surnames dataset, available at \url{https://www.census.gov/data/developers/data-sets/surnames.html}.
This dataset contains 162k surnames of US residents along with a breakdown of the fraction of people with this surname of race categories: ``white'', ``black'', ``asian/Pacific islander'', ``American Indian'', ``mixed race'' and ``hispanic''.
The advantage of using this data set is the amount of data it contains and its known high accuracy.
Since the US has has many immigrants, the census data should have a high diversity of surnames.
In order to focus on categories most likely to yield successful training, we restrict this to the four categories of ``white'', ``black'', ``Asian'' and ``hispanic''.

We prepare the surnames for input into the DNN by converting each into a vector of reals in the range $[0,1]$ with uniform length, padded to the left to width 20 by zeros.
Each letter is converted according to its position in the alphabet, meaning $\textrm{A}\to 0$ and $\textrm{Z}\to 1$.
The output target for each surname is the fraction of people with that surname belonging to each of the four race categories, so a vector of length 4 constrained to the range $[0,1]$.
We create a model using 4 dense layers with widths 200, 64, 32 and 4.
The model is trained using a mean-squared loss function and the Adadelta optimizer over 300 epochs.

While using the full data set, we saw the model become trapped in an over-training condition where all surname inputs yielded the same output: ``white'';
the model had learned that most people in the training set were white and just assumed everybody was.
Although this has humorous parallels to our actual society and the problem of racism, it did not yield a particularly useful model for our purposes.
We corrected the training set by dividing it up according to which race was most common for each surname, then selecting only 7k names from each of the four subsets, totalling 28k.

After training on this restricted set, the model yielded results, but with poor accuracy.
We decided to reorganize the input data so that the real input corresponded to the frequency of the letter rather than its ordinal position in the alphabet;
that is $\textrm{E}\to 0$, followed by T, A, etc.
This produced only modest gains.

Next, we expanded the input data such that rather than sharing a single real input, each letter would be represented by a bit in a vector with length covering the alphabet.
That is, $\textrm{E}\to \{1, 0, \dots, 0\}$ and $\textrm{Z}\to \{0, 0, \dots, 1\}$.
This increases the dimension of the input to 520, greatly increasing the number of model parameters and resulting in a model approaching usable.

The test data set is the surnames of all lattice-paper authors, sorted by the number of papers written.
This model worked reasonably well: it recognised ``AOKI'', ``LIU'' and ''HASHIMOTO'' as ``Asian'', and  ``JANSEN'', ``LEINWEBER'' and ``HELLER'' as ``white''.
However, there were a number of issues:
It predicts ``LIN'' as ``white'', possibly due to the fact that ``-LIN'' is a common ending in surnames such as ``FRANKLIN'').
This model also tagged far more authors in this data as ``black'' than our field has.
This can be understood due to American history;
black and white Americans simply share many surnames, and there is a mixture of surnames in the training data.
For example, ``SMITH'' is about 50\% white and black according to the census data.
These American trends do not well match the demographics of the lattice authors data set.

At this point, we decided to pivot away from the census training set.
Rather than attempting to discern race from surnames (it seems likely that including first names would be necessary to even attempt this), we would attempt to identify the ethnic origin of the surnames.
By training a network with the most common surnames in a variety of diverse countries, we would be able to better cover the lattice data.
The new training data included the most common 1000 surnames from 27 countries: Spain, Mexico, Brazil, Argentina, Italy, France, England, Scotland, Ireland, Germany, Greece, Sweden, Russia, Poland, Hungary, Turkey, Israel, Egypt, Syria, Iran, Pakistan, India, Vietnam, China, Taiwan, Japan and South Korea.
The new training set included Unicode characters, unlike the census data, and we decided to retain them, hoping they would be strong indicators of ethnicity.
This expanded our alphabet to 42 characters, and the padded length to 21, yielding an input dimension of 882.

We adjusted the model to four layers with widths 1764, 441, 216 and 27, giving the final layer a sigmoid activation to constrain it to $[0,1]$.
The output is expected to be a unit vector, but we left it unnormalized.
Since the new model is a kind of categorical assignment, we changed the loss function to categorical cross-entropy and the optimizer to Adam, which is expected to converge more smoothly for this kind of model.
The new model has 2.4M parameters.

With a few parameters tuning, we got about 3\% errors on the overall training.
We then have the network to print the top-3 prediction for the country of the last names.
Many previously problematic surnames were resolved;
it identified ``JANSEN'' as	Swedish,
``HELLER'' as German, and
``LIN'' as Taiwanese.
The model predicted ``LEINWEBER'' was French, but German as a second guess.
We hand-checked the top-323 surnames, covering all authors with 20 or more lattice papers, and found 77\% accuracy at the country level and 92\% accuracy discerning white versus Asian surnames.
We use the hand-corrected 323 surnames plus the ML-evaluated 2500 remaining surnames in the remainder of this analysis.
To simplify the analysis, we broadly categorize authors as either the field-dominating ethnicity, ``white'', or not;
we presently lack the statistics and precision necessary for finer subdivision~\cite{Aubin:2019rdf}.

\begin{figure}[tb]
\centering
\includegraphics[width=0.32\textwidth]{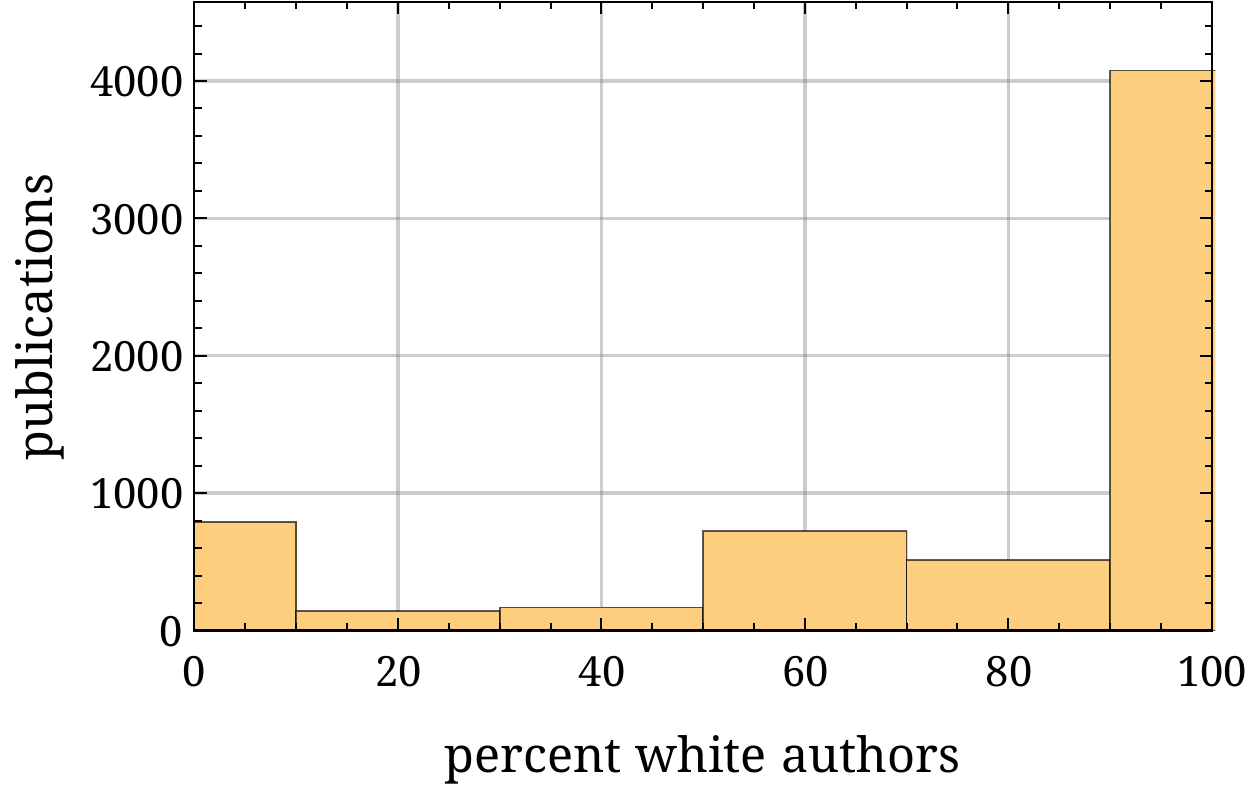}
\includegraphics[width=0.32\textwidth]{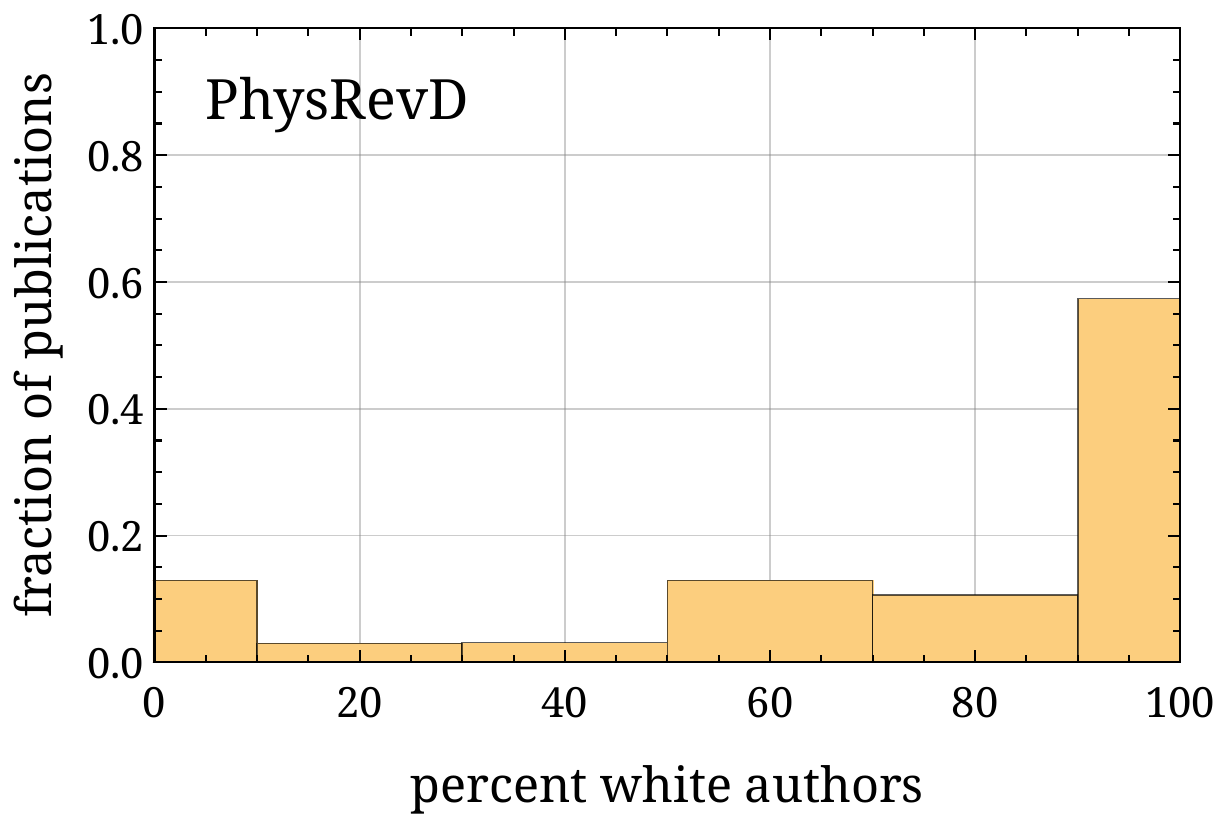}
\includegraphics[width=0.32\textwidth]{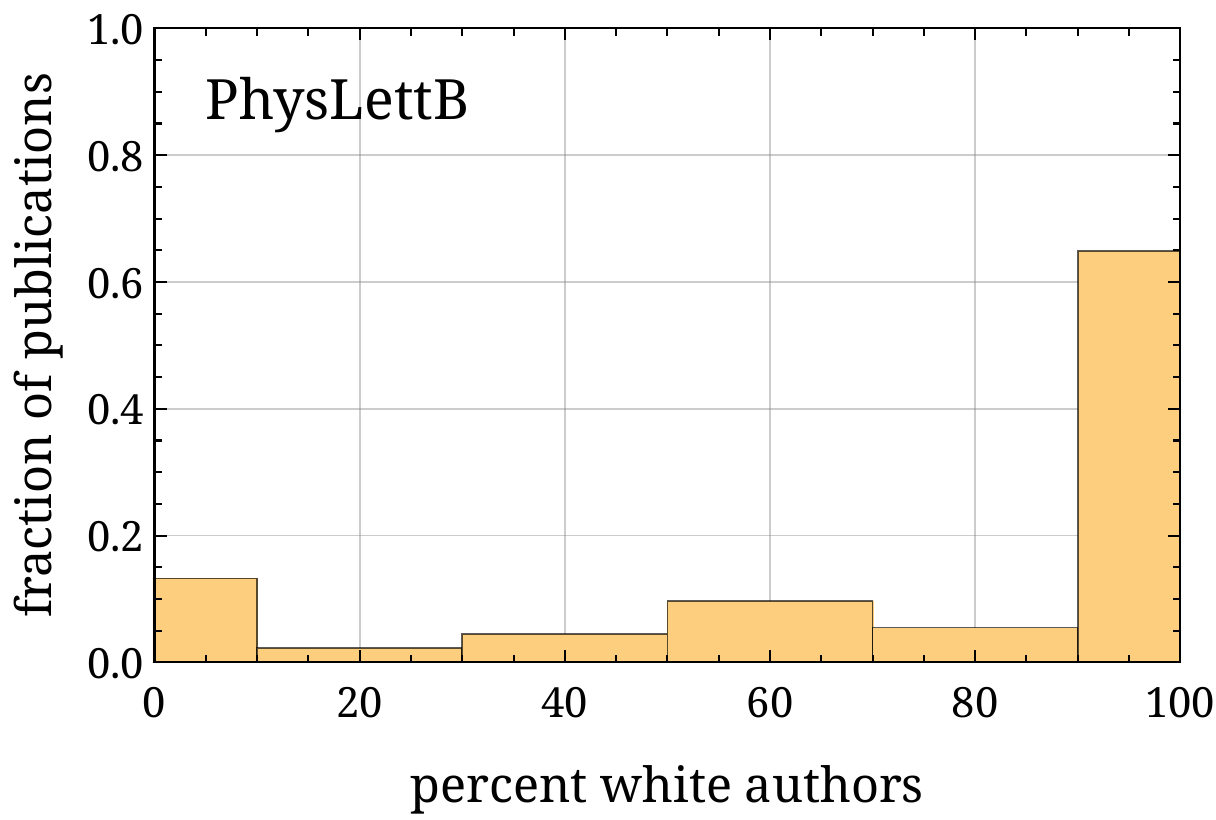}
\includegraphics[width=0.32\textwidth]{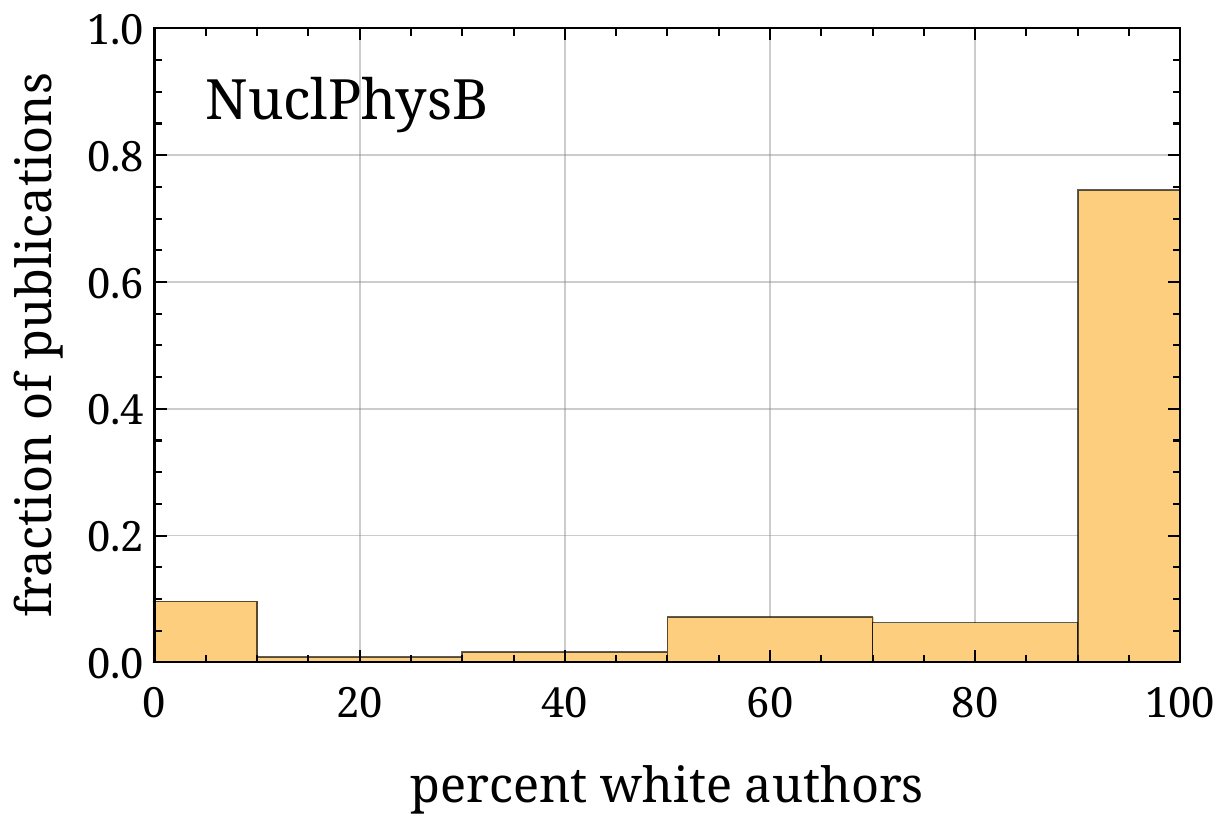}
\includegraphics[width=0.32\textwidth]{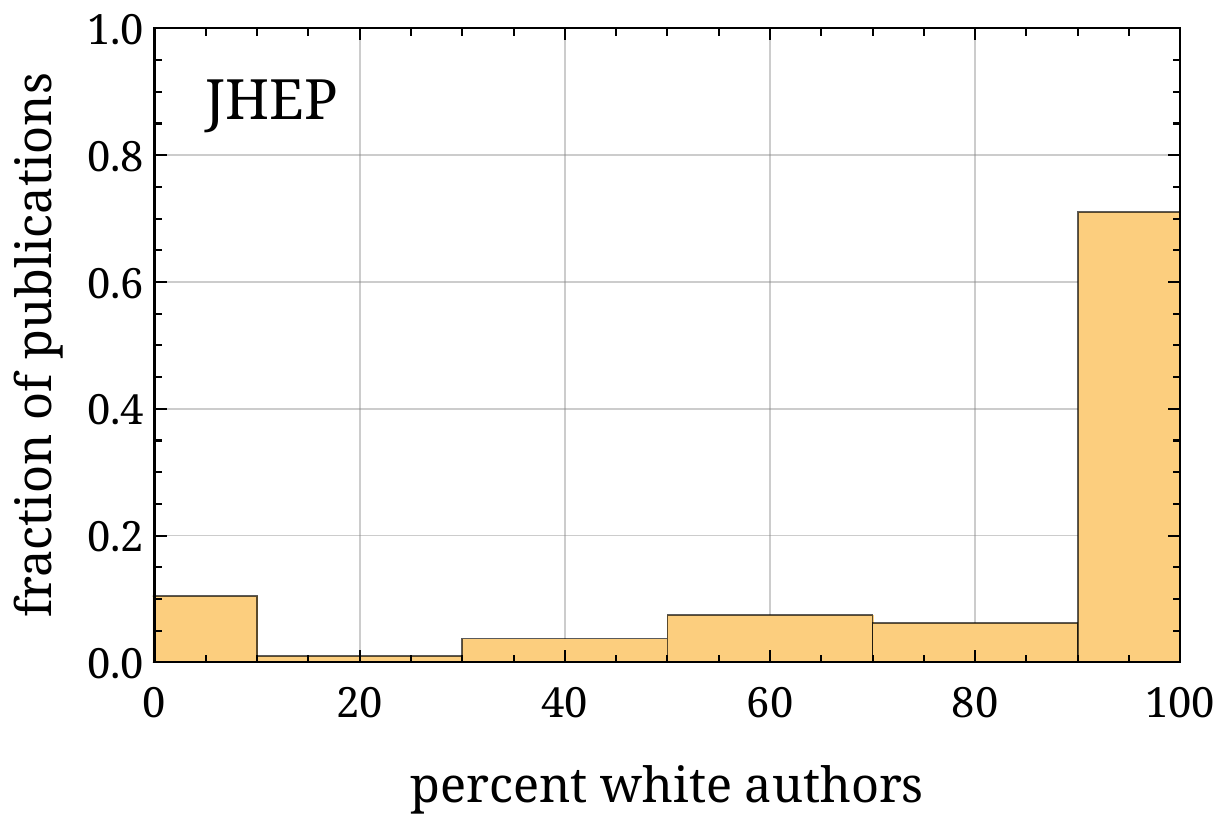}
\includegraphics[width=0.32\textwidth]{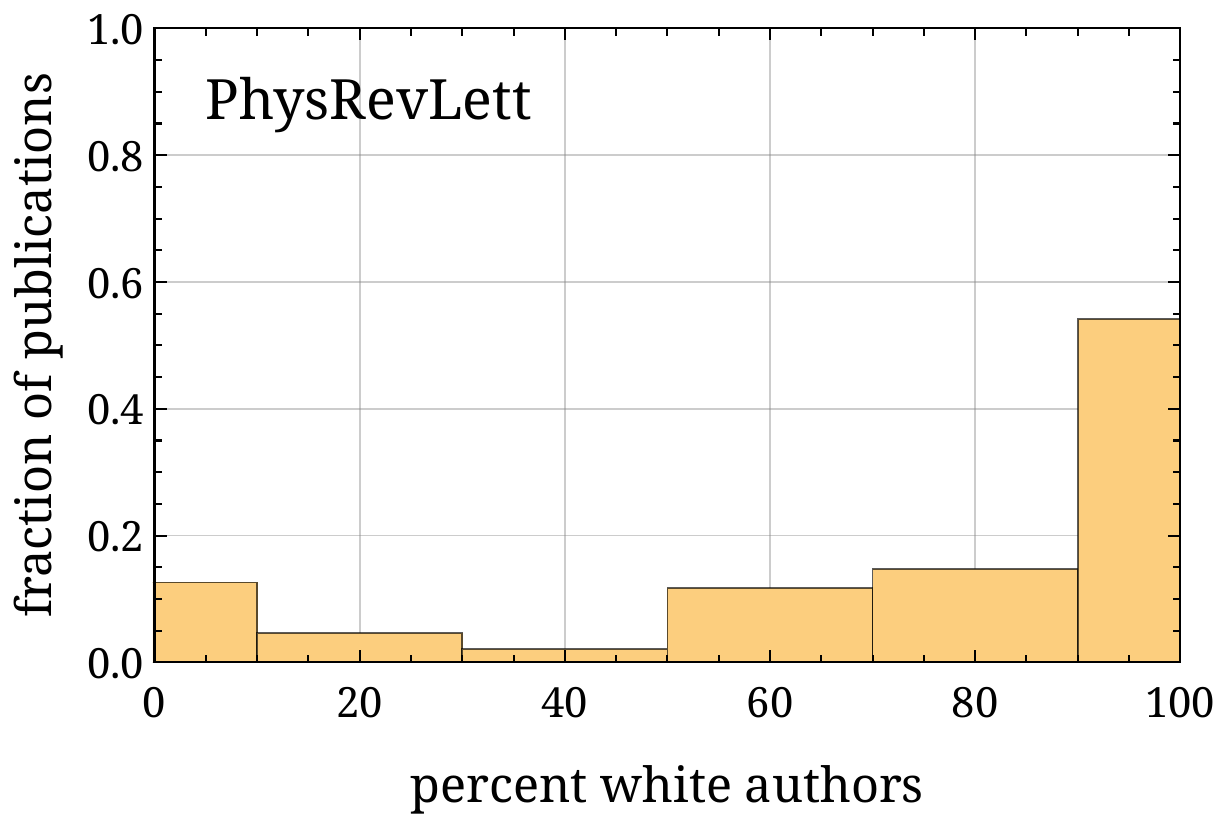}
\caption{
Histograms of the author-list composition for all published papers recorded on arXiv (upper left) and the breakdowns for the top-five journals for lattice publication.
\label{fig:ethnicity-hist}}
\end{figure}

We first look at the total number of papers published versus the fraction of authors identified as white in the author list.
The upper-left part of Fig.~\ref{fig:ethnicity-hist} shows that 81\% of publications have 50+\%  white authors. 
For comparison, the 2019 climate survey showed ``Caucasian/white'' to be about 75\% of those responding (see  Fig.~8 in Ref.~\cite{Aubin:2019rdf}).
We then look into the splits of the data for the top-five lattice-paper journals; 
they are    Phys. Rev. D,       Phys. Letter B,   Nucl. Phys. B,    JHEP,   and Phys. Rev. Lett.
The fraction of publications with majority-white author lists is    81\%,               86\%,             91\%,             85\%,       80\%, respectively.
Once again, none of the above results include those the journal declined to publish.

\begin{figure}[tb]
\centering 
\includegraphics[width=0.5\textwidth]{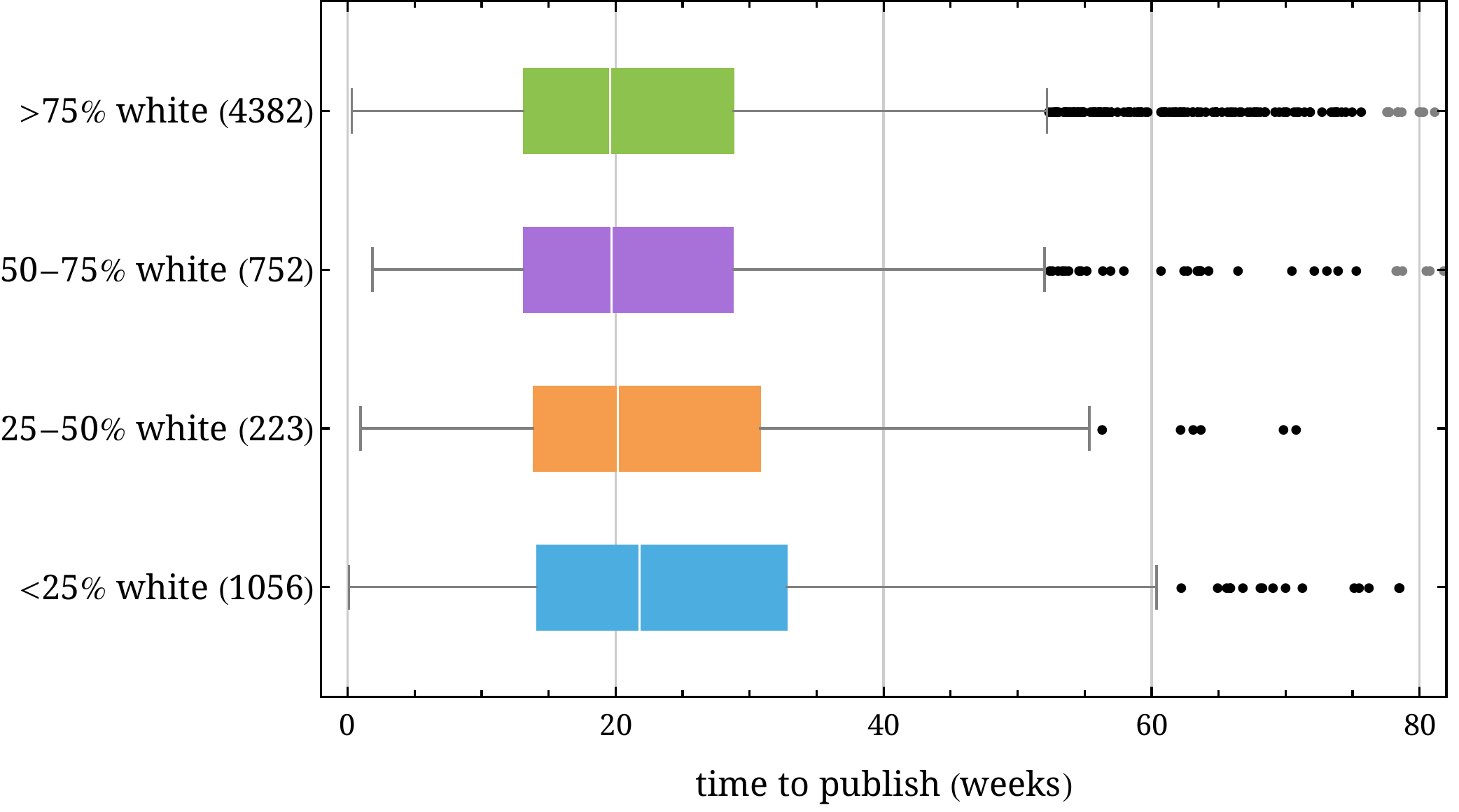}
\includegraphics[width=0.4\textwidth]{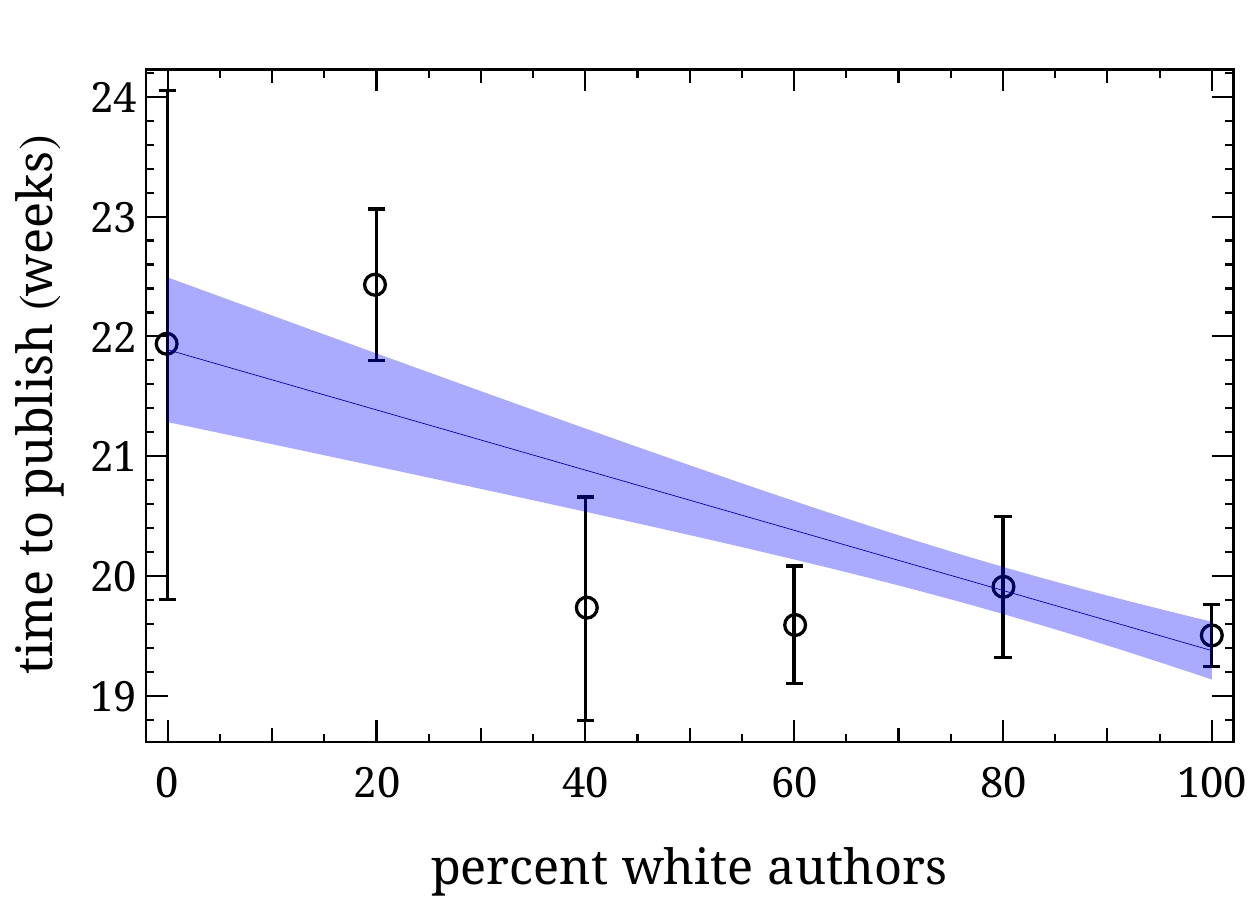}
\caption{
Time to publication as a function of author ethnicity fraction (left) as box-and-whisker plots and (right) fit to a linear trend.
There is evidence that all or mostly white author lists see faster times to publication.
\label{fig:ethnicity-publishtime}}
\end{figure}

Figure~\ref{fig:ethnicity-publishtime} shows how time to publication varies as a function of ethnicity distribution.
On the left, the box-and-whisker plots show the median and quartiles for author-list compositions binned into four groups.
We see that there is an increase of about 10\% in the median for papers with few white authors.
On the right-hand side, we bin the groups to the nearest 20\%, and plot the median and standard deviation of the time to publication as the black points.
We fit this to a linear trend, finding a best fit $19.4(2) + 2.5(7)(1-p)$;
that is, an all-nonwhite authored paper expects to be tied up in the editorial process for 2.5 weeks longer than an all-white author group.

\section{Author Gender Study}

Having had some success with using machine learning on author ethnicity, we attempt to study author gender using machine learning.
Prior to beginning this phase, we thought machine learning would work well for this application, at least for Western first names.
Many Western languages have grammatical gender that makes names have indicative endings (e.g. the -o/-a rule), and we should be able to train the network to use such features to give us the gender of the authors.
In practice, there are many problems.
Despite the strong incentive for authors to ensure that they are credited for their work by having their correct full name attached to it, the paper metadata are full of typos, abbreviations and inconsistencies.
Thus, a lot of effort is needed to clean up the data set.
We use inSpire's unique author identifier, when available, to determine an author's ``preferred name''.
If a preferred name is not available, the ``identifier name'' is used instead.
If no author identifier is present, we fall back on whatever name is printed in the paper name.
Since last names are always available (though sometimes inconsistent), we group names with identical last names and matching initials, checking for variant spellings such as omitted accents or hyphens.
We replace all names in the group with a matching preferred (or identifier) name;
it is possible that this step introduces error if an author without a unique author identifier has the same initials as another author who has one.

Using the lessons learned from the ethnicity study, we trained a neural network on the most common first names from various countries, and test a random sample of 100 lattice-author names to see how well the network works.
The training score achieved 90\%, but the network predicts higher female fractions than male, by 30\% more.
Due to the diverse backgrounds of lattice authors, the network has trouble finding consistent rules across the data set.
It is possible that including last names, which were a good indicator of ethnicity might help the network correctly identify gender.
However, at this stage, we abandon the network in favor of a simpler lookup-table scheme.

With the failure with the machine-learning approach, we still needed an efficient way to resolve 3109 authors in the arXiv and inSpire database using as little human intervention as possible.
We took advantage of the past few years of the lattice-conference participant lists (2019--2014) and the plenary-speaker list since 2000.
Most lattice authors will have attended a lattice conference during this period, and older papers will have senior authors who will have given a plenary. 
A database derived from lattice participant lists allowed us to identify 714 authors, leaving 2395 remaining unidentified.
To bring the number down, we sorted the remaining authors according to the number of papers authored, taking those with 20 or more papers and making private inquiry to colleagues who personally knew those authors.
For the remaining 2153 authors, we created a lookup table from the neural-network training set, resolving a further 1296.
The final 862 unresolved authors (most having only an initial listed in the database) are simply assumed to be male;
these authors had an average of 2 papers, and none had more than 8 papers.
We also discarded 12 papers that only list a collaboration name.

\begin{figure}[tb]
\centering
\includegraphics[width=0.45\textwidth]{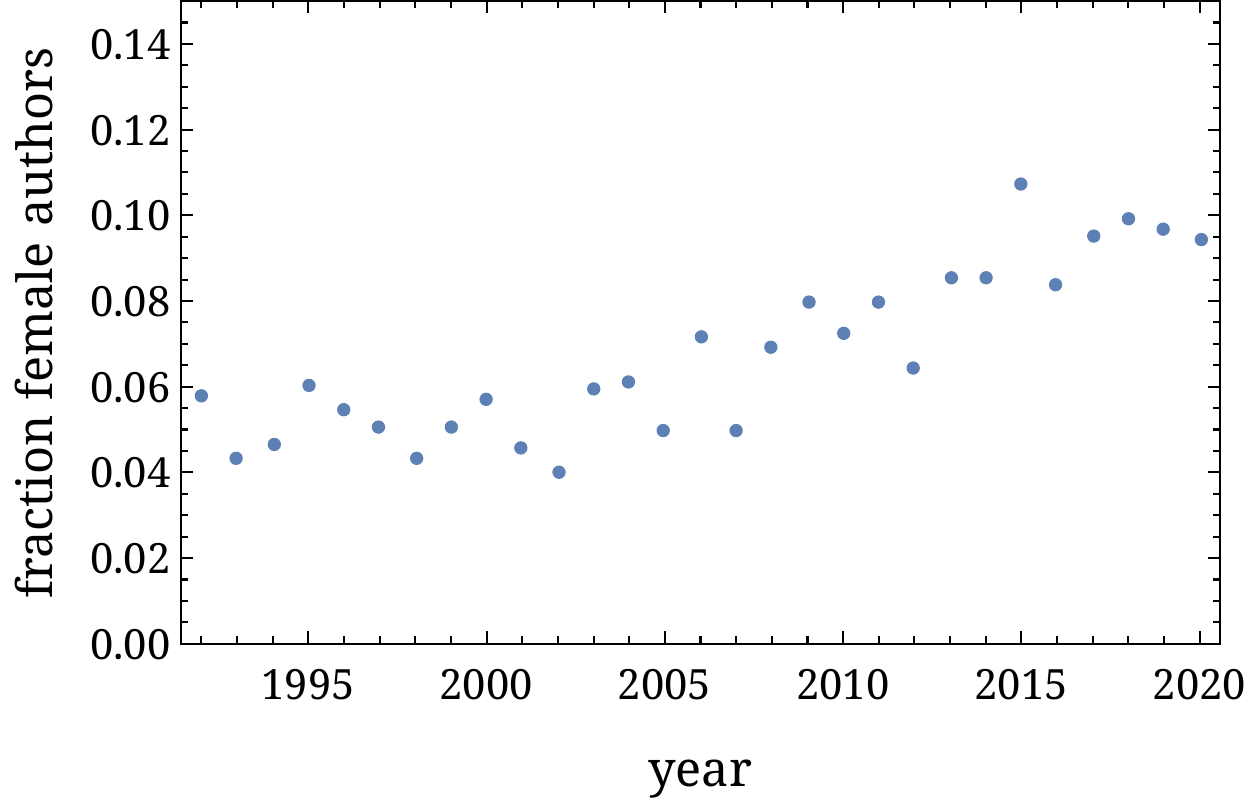}
\includegraphics[width=0.45\textwidth]{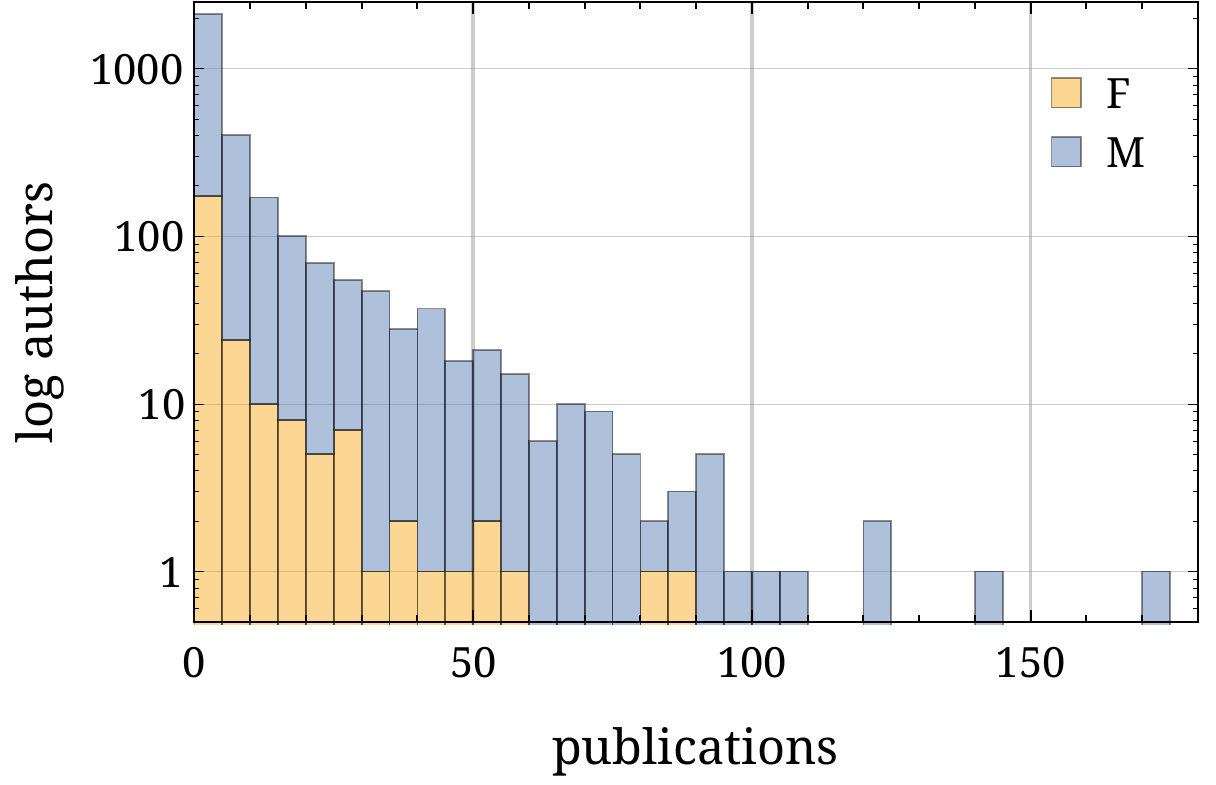}
\caption{
(Left) The fraction of female authors on the lattice papers as function of year. (Right) The log histogram plot of the number of male (blue) and female (yellow) authors as a function of published hep-lat papers for that author.
The publication numbers are binned by 5.
The number of authors decreases exponentially with number of publications, but the drop in female authors is more steep than male authors.
\label{fig:gender-author-info}}
\end{figure}

Using the fully assigned data set, we can study gender differences in paper publication.
The left-hand side of Fig.~\ref{fig:gender-author-info} shows the fraction of female authors in primary hep-lat peer-reviewed published papers as a function of year.
There is a slow increasing trend, which is encouraging.
On the other hand, extrapolating suggests that it will take another two decades for the women author fraction to match the AIP 2017'S 20\% of female physics doctorates~\cite{AIP2019}.

The right-hand side of Fig.~\ref{fig:gender-author-info} shows a log histogram of male and female authors as a function of the number of publications by that author.
It is perhaps not surprising to see that the majority of lattice authors publish fewer than 5 papers.
The first bin is likely undergraduate and graduate students who worked briefly in the field, published a few papers for their thesis work, and left the field;
there is an 11:1 male to female ratio, which is close to  the past female fraction of conference participants~\cite{Lin:2016age}.
The most prolific female author has 85 papers, while the top male author has 173.
Between these extremes, there is a nearly exponential drop in the number of authors as a function of the number of publications.
There appears to be a sharper drop in female publications relative to male, which may reflect the pipeline problem;
this is concerning. 

\begin{figure}[tb]
\centering
\includegraphics[width=0.32\textwidth]{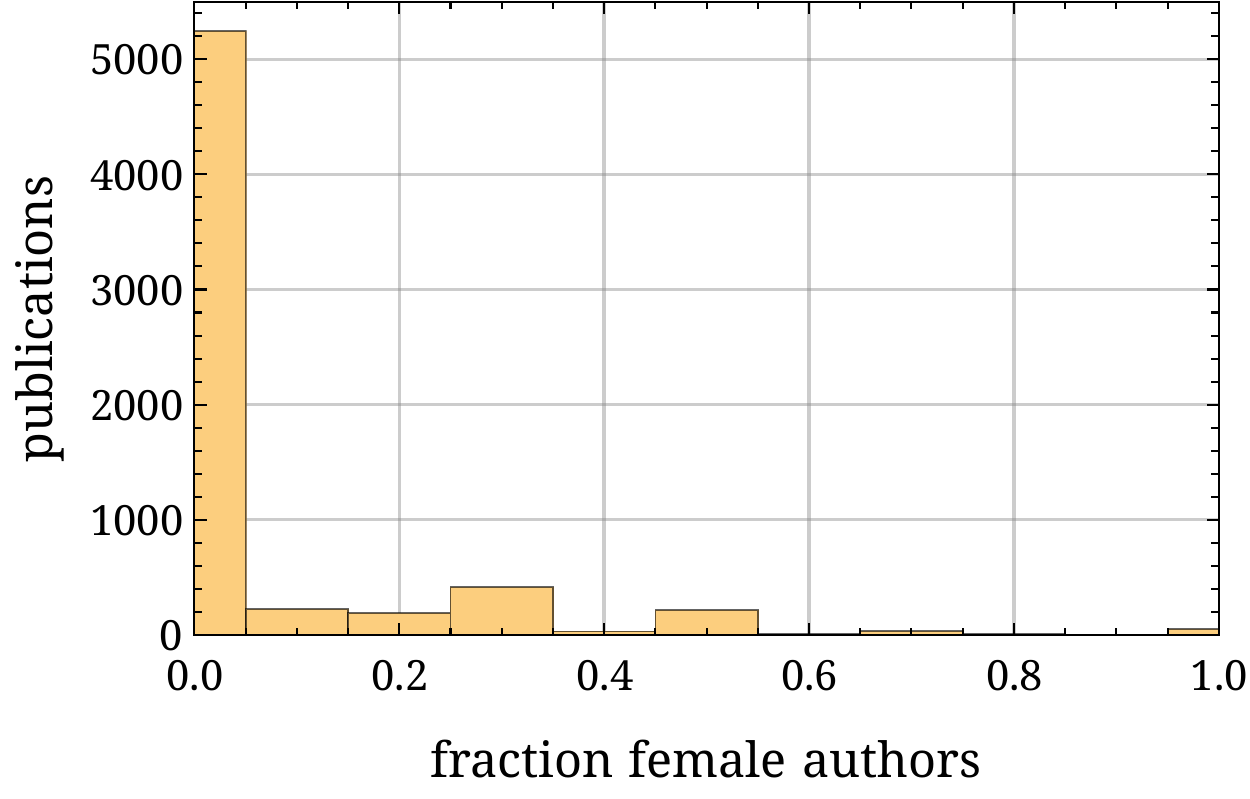}
\includegraphics[width=0.32\textwidth]{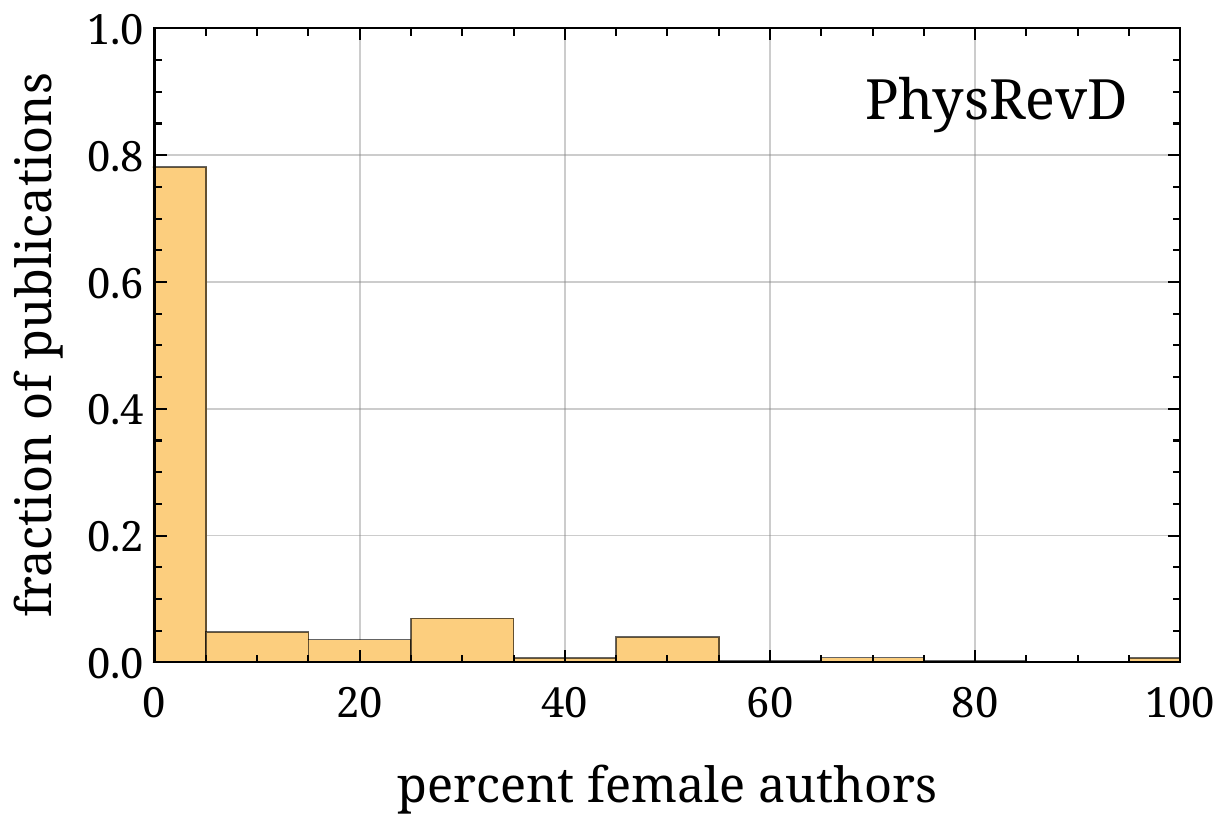}
\includegraphics[width=0.32\textwidth]{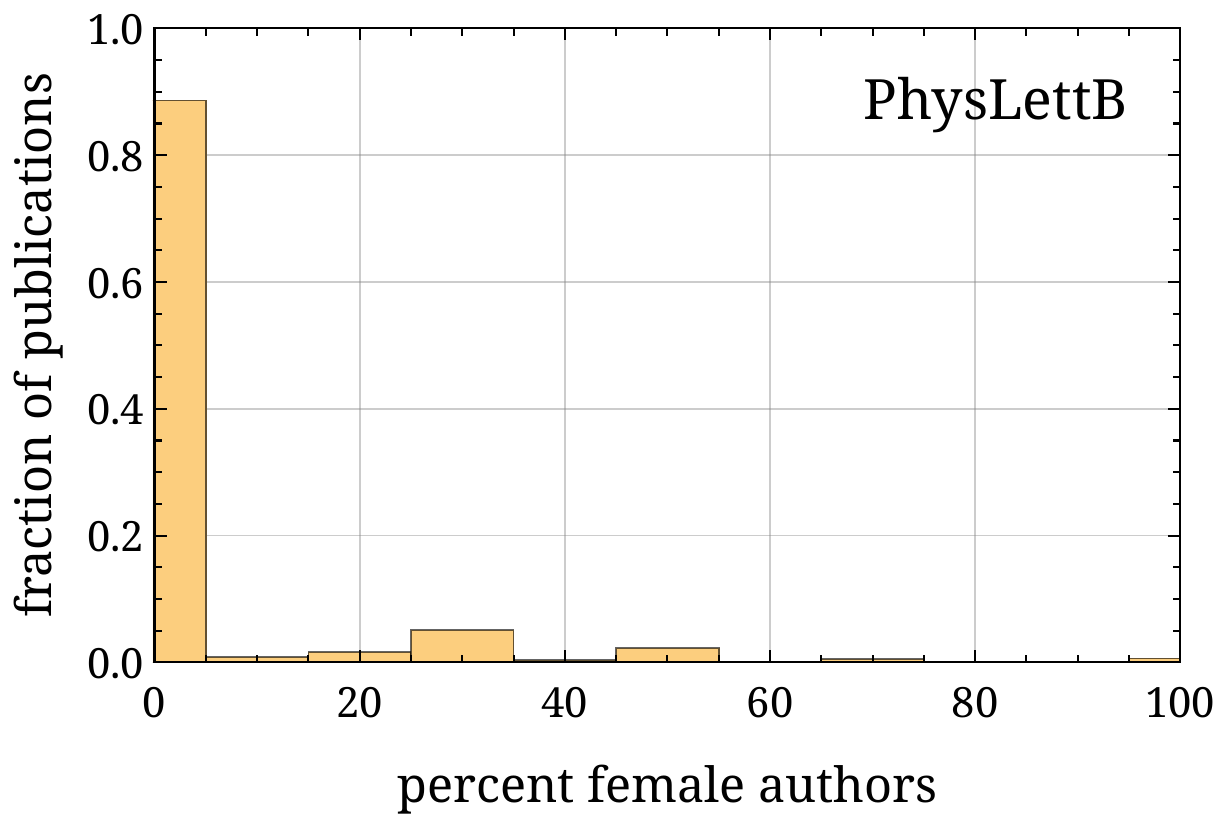}
\includegraphics[width=0.32\textwidth]{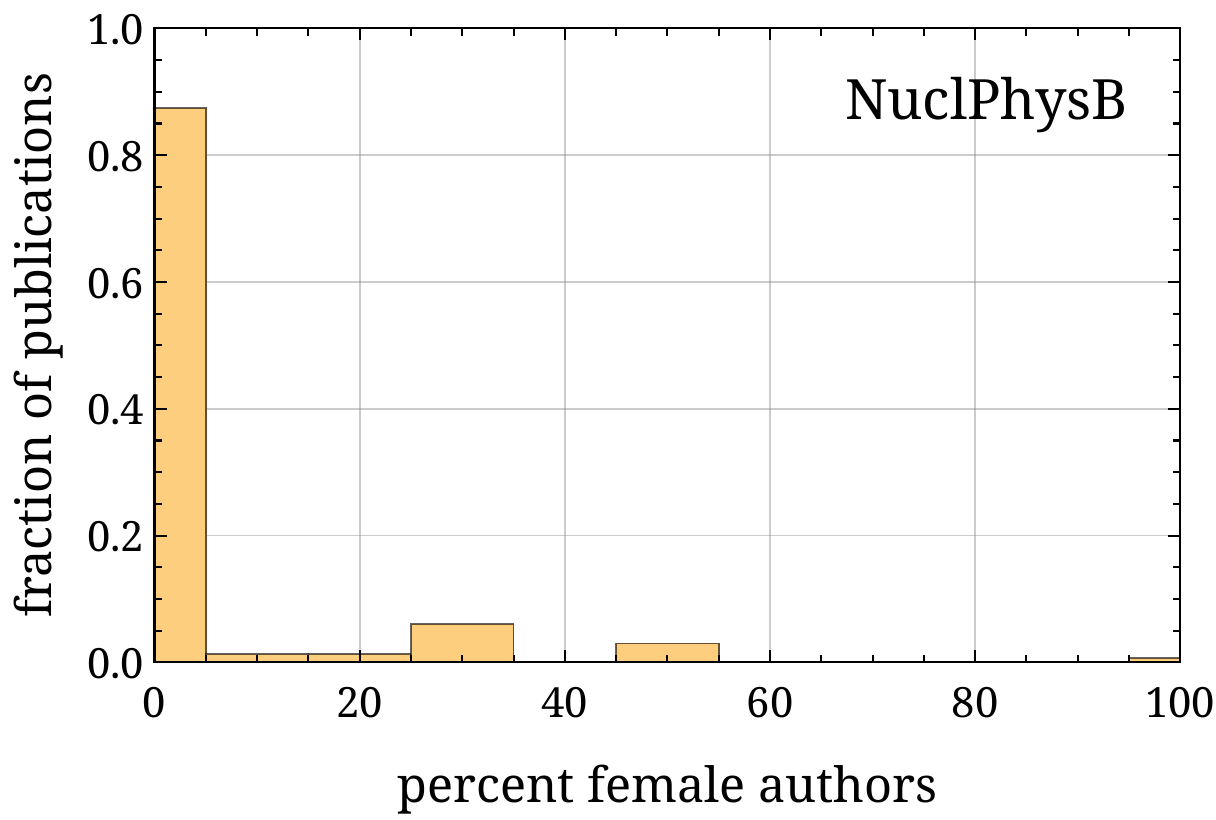}
\includegraphics[width=0.32\textwidth]{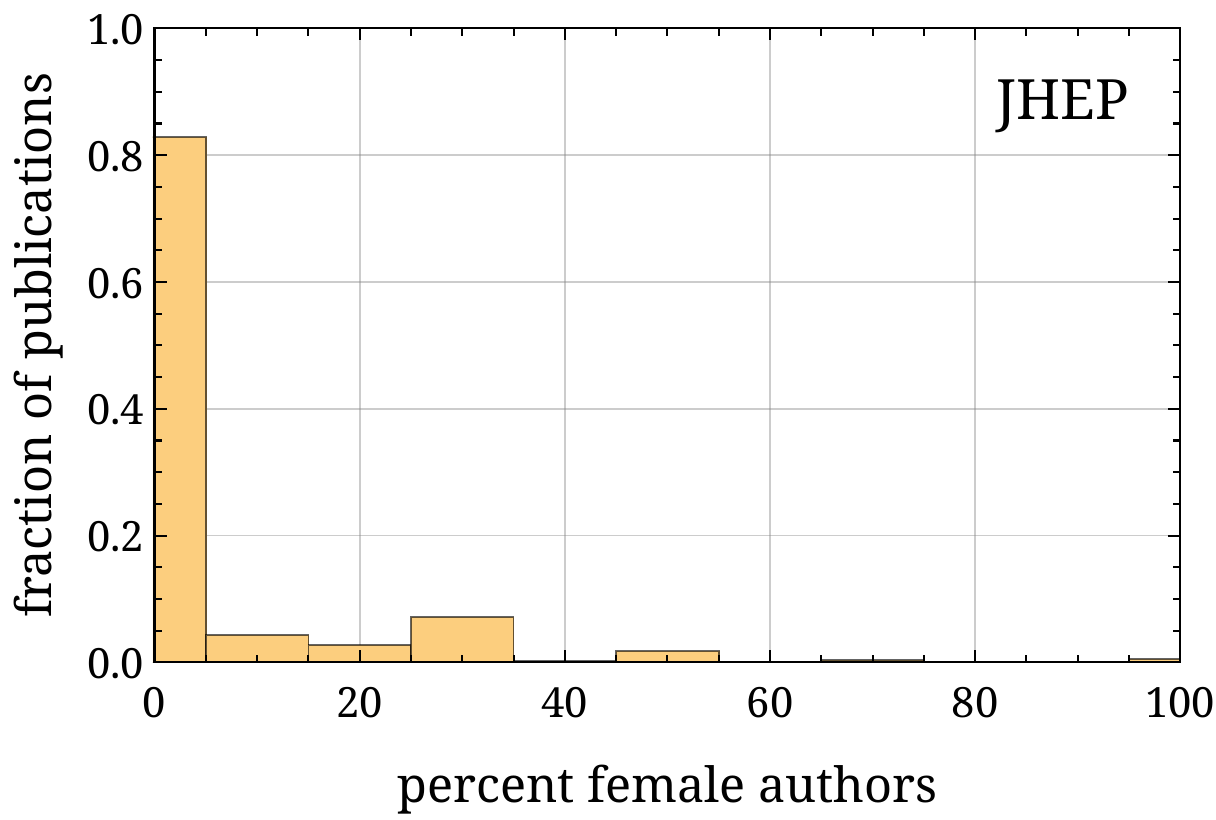}
\includegraphics[width=0.32\textwidth]{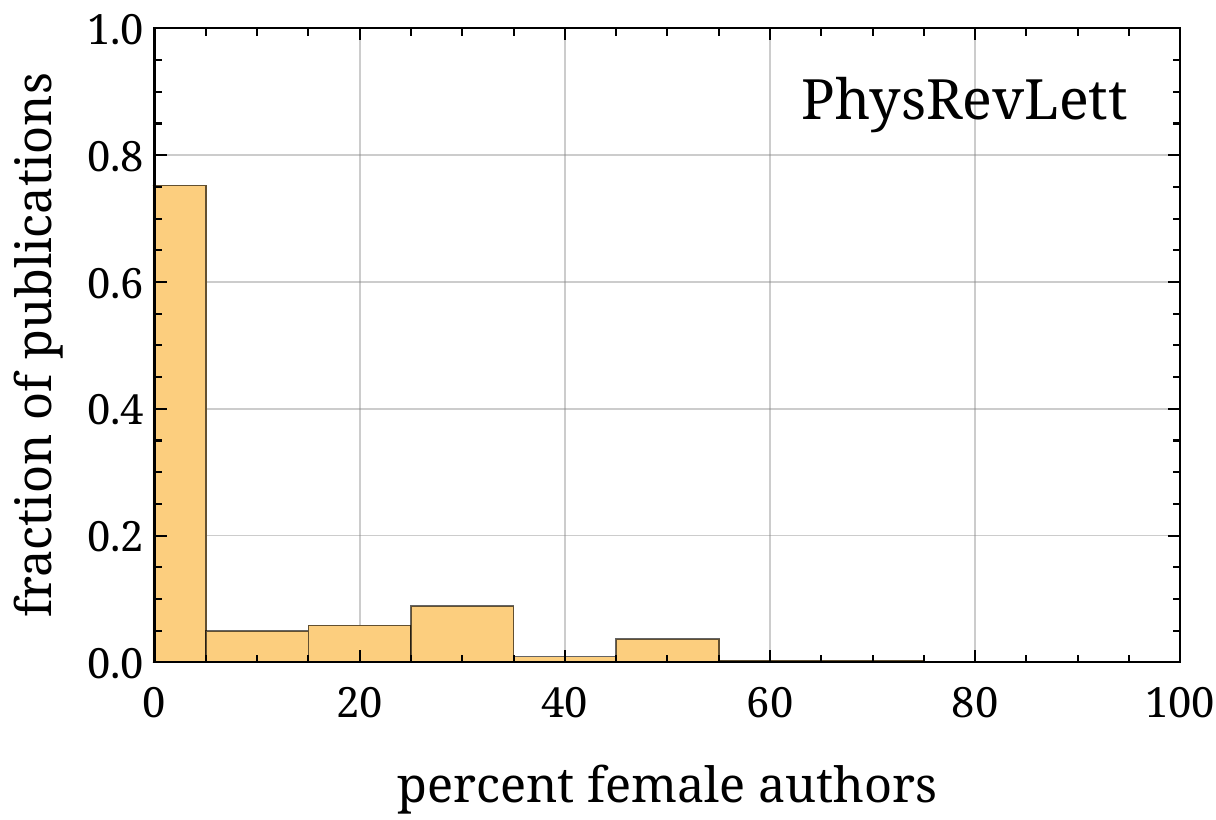}
\caption{
Histograms of the author-list composition for all published papers recorded on arXiv (upper left) and the breakdowns for the top-five journals for lattice publications. 
\label{fig:gender-hist}}
\end{figure}

We then look at the histogram of papers as a function of the fraction of female authors on each publication's author list in Fig.~\ref{fig:gender-hist}.
As expected, most of the lattice papers are published with 5\% or fewer female authors.
There are small peaks associated with women being 1 in 2 or 3 authors on a paper, but very few papers with more women than men.
We further break down the papers from the top journals, binning by 20\% due to the smaller statistics.
There are 10\% or fewer papers with more than 20\% female authors;
some journals, such as \textit{Phys. Rev. Lett.} and \textit{Comput. Phys. Commun.}, have never published a lattice paper without a male author on it.

\begin{figure}[tb]
\centering 
\includegraphics[width=0.4\textwidth]{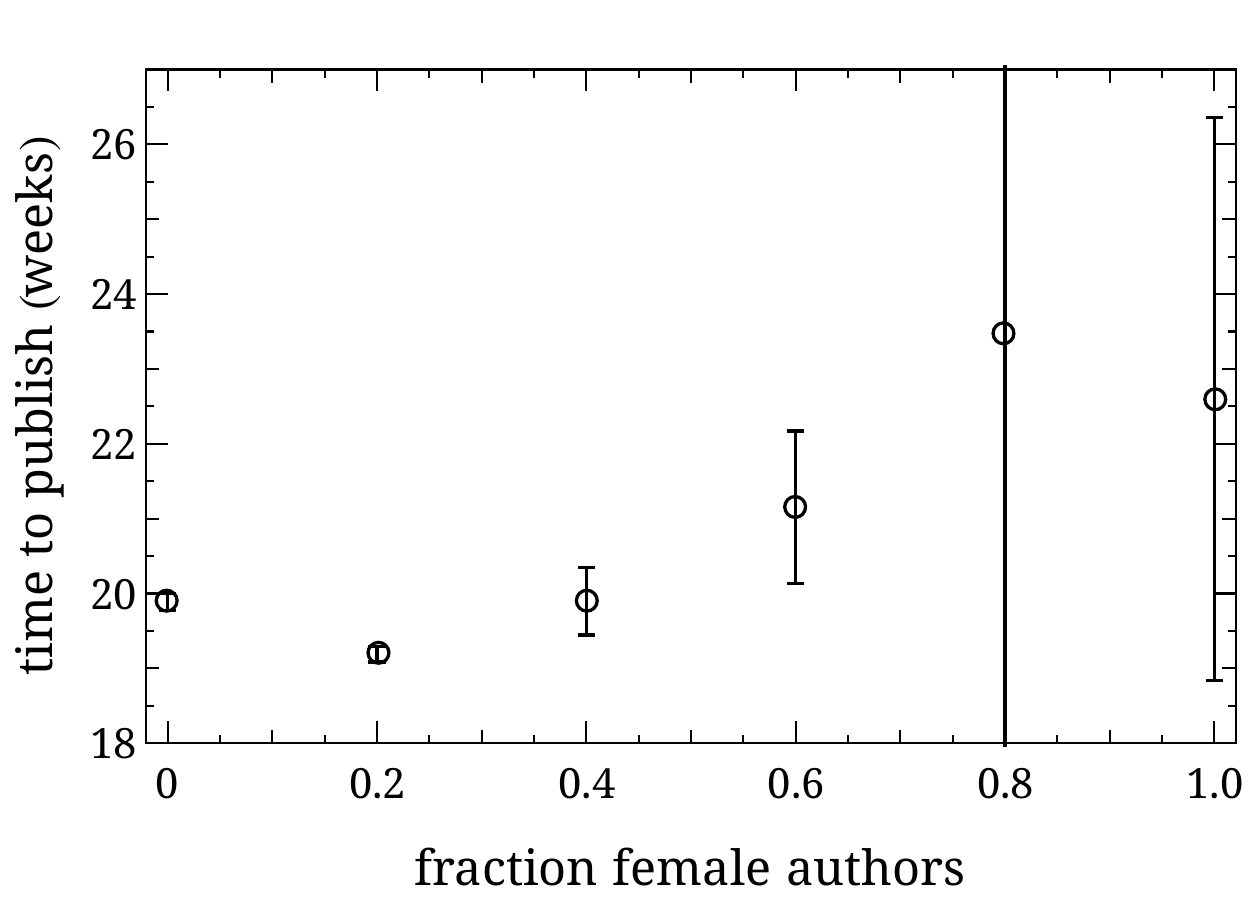}
\includegraphics[width=0.45\textwidth]{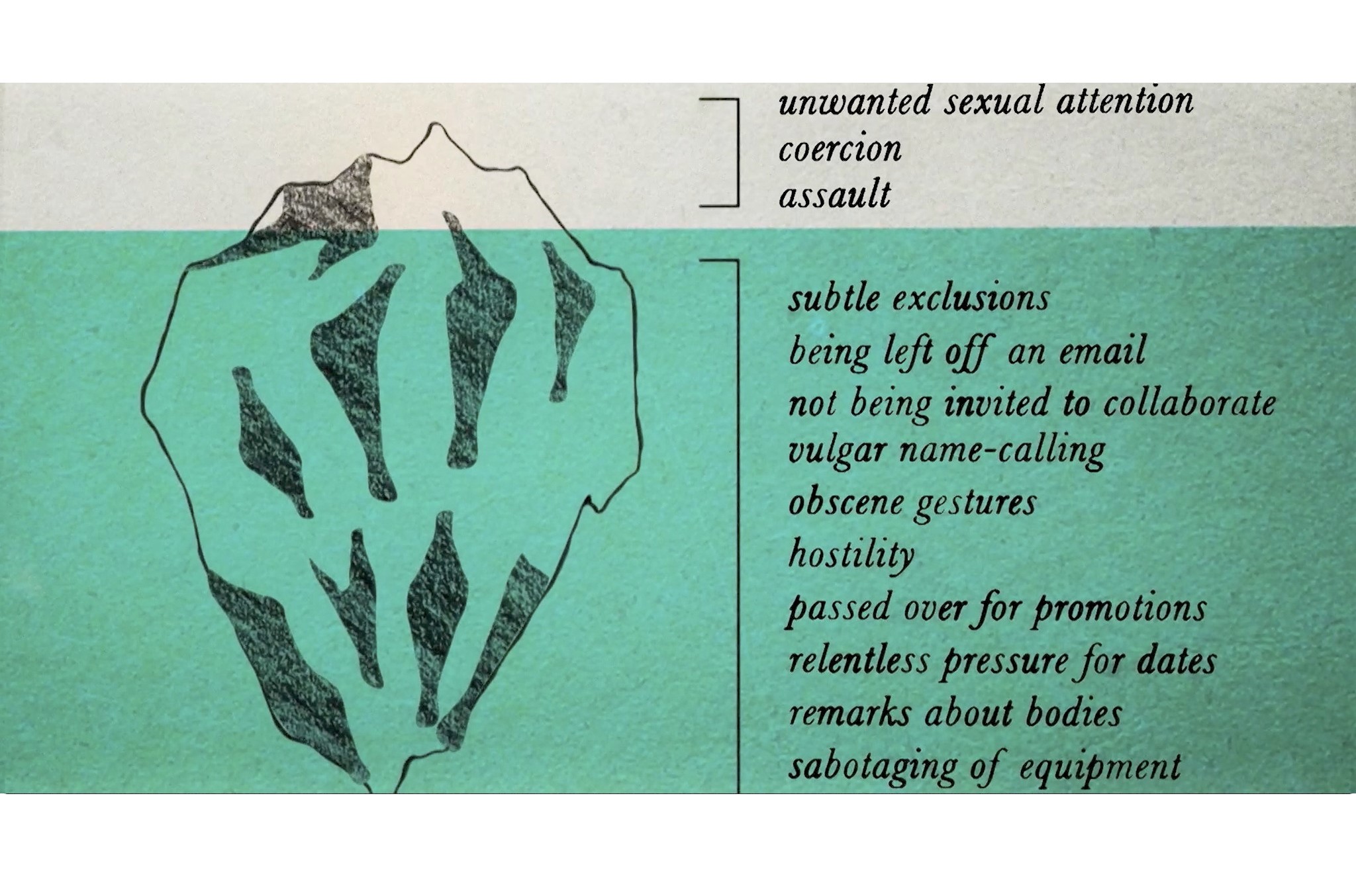}
\caption{
(Left) Time to publication as a function of author gender fraction.
(Right) The list of under-the-iceberg obstacles female scientists often encounter from the documentary film ``Picture a Scientist''~\cite{PictureAScientist}
\label{fig:gender-publishtime}}
\end{figure}

Figure~\ref{fig:gender-publishtime} shows how time to publication varies as a function of gender distribution. 
We bin the groups to the nearest 20\%, and plot the median and its standard deviation of the time to publication as the black points. 
There appears to be an increase in publication time for papers by primarily female authors.

\section{Conclusions}

We studied the hep-lat peer-reviewed published papers, examining ethnicity and gender of authors, and time to publication.
We performed analysis for papers published in all journals and also broke out the top-5 journals in terms of number of lattice papers published.
We see some evidence of trends in time to publication versus author ethnicity and the fraction of female authors.
All these data reflect a period when peer review is a common practice in the field;
most reviews are one-sided communications, and report quality differs greatly according to the reviewers' available time, background and agenda.
Conscious and unconscious bias in referee reports may be significant, but since none of these reports become part of the public record, it is difficult to ascertain whether there is a gender and/or ethnicity bias in referee reports.
Similar work submitted around the same time can experience very different review processes, leading to delays or different editorial decisions.
Since our study can only access the records of published papers, we do not have information on those papers by minority or female authors that did not make it to publication.
In any scientific process, we must be wary of the encroachment of human bias influencing results.
This should be true of the journal editorial decision making, since the papers published in prestigious journals can sometimes shape the research directions.
Some journals are now taking steps to improve the traditional peer-review process outside physics~\cite{NaturePeerReview}.

According to Ref.~\cite{RacialCitation}, an article in \textit{Inside Higher Ed}, ``higher education, despite some representational inroads, remains a white institutional space, with highly racialized patterns of access, resources and rewards.'' 
For female scientists, the documentary ``Picture a Scientist''~\cite{PictureAScientist}, lists the difficulties encountered even for those who are successful, as shown on the right-hand side of Fig.~\ref{fig:gender-publishtime}.
With the ongoing COVID, a report in \textit{Nature}~\cite{NatureCovidImpact} states: ``A survey of principal investigators indicates that female scientists, those in the `bench sciences' and, especially, scientists with young children experienced a substantial decline in time devoted to research. This could have important short- and longer-term effects on their careers, which institution leaders and funders need to address carefully.''
These trends all indicate that action is needed to ensure a healthy and diverse scientific environment for the future of our beloved research field.

\section*{Acknowledgments}
We thank Etsuko Itou, Meifeng Lin, Liuming Liu, Shigemi Ohta, Andreas Kronfeld, Pavel Nadolsky and Saul Cohen for their help in identifying authors with first names missing from the metadata, or identifying the genders of first names. 

\input{main.bbl}
\end{document}

%% file: main.bbl
\providecommand{\href}[2]{#2}\begingroup\raggedright\endgroup